\documentclass[preprint,5p,twocolumn]{elsarticle}
\usepackage{amsmath}
\usepackage{hepunits}
\usepackage{multirow}
\usepackage{gensymb}
\usepackage{hyperref}
\usepackage{float}
\DeclareSIUnit\weight{wt}
\usepackage{siunitx}
\usepackage{graphicx}
\usepackage[dvipsnames]{xcolor}
\usepackage{tikz}
\usepackage[normalem]{ulem}
\usepackage{ifthen}
\usepackage{textcomp}

\usetikzlibrary{arrows}
\usetikzlibrary{positioning}

\journal{Nuclear Instruments and Methods in Physics Research~A}

\hypersetup{
  pdfstartview={XYZ null null 1},
  pdfauthor={J.M. Gómez-Guzmán, K. Bernert, A. Devishvili, C. Klauser, B. Märkisch, U. Schmidt, T. Soldner},
  pdftitle={Depolarization studies on low-depolarizing Cu/Ti and Ni(Mo)/Ti neutron supermirrors}
}
\begin{document}

\title{Depolarization studies on low-depolarizing Cu/Ti and Ni(Mo)/Ti neutron supermirrors}

\author[1]{Jose Manuel Gómez-Guzmán\corref{cor1}}
\ead{jose.gomez@frm2.tum.de}
\author[2]{Karina Bernert}
\ead{karina.bernert@tum.de}
\author[3]{Anton Devishvili}
\ead{devishvili@ill.fr}
\author[4]{Christine Klauser}
\ead{christine.klauser@psi.ch}
\author[2]{Bastian Märkisch}
\ead{maerkisch@ph.tum.de}
\author[5]{Ulrich Schmidt}
\ead{ulrich.schmidt@physi.uni-heidelberg.de}
\author[3]{Torsten Soldner}
\ead{soldner@ill.fr}
\cortext[cor1]{Corresponding author}

\address[1]{Heinz Maier-Leibnitz Zentrum, Technische Universität München, 85748 Garching, Germany}
\address[2]{Technische Universität München, James-Franck-Str. 1, 85748 Garching, Germany}
\address[3]{Institut Laue-Langevin, 71 avenue des Martyrs, CS 20156, 38042 Grenoble Cedex 9, France}
\address[4]{PSI Center for Neutron and Muon Sciences, Forschungsstrasse 111, 5232 Villigen PSI, Switzerland}
\address[5]{Physikalisches Institut, Universität Heidelberg, Im~Neuenheimer~Feld~226, 69120 Heidelberg, Germany}

\begin{abstract}
  Neutron supermirrors are a crucial part of many scattering and particle physics experiments. So far, Ni(Mo)/Ti supermirrors have been used in experiments that require to transport a polarized neutron beam due to their lower saturation magnetization compared to Ni/Ti supermirrors. However, next generation $\beta$ decay experiments require supermirrors that depolarize below $\num{e-4}$ per reflection to reach their targeted precision. The depolarization of a polarized neutron beam due to reflection from Ni(Mo)/Ti supermirrors has not yet been measured to that precision. Recently, Cu/Ti supermirrors with a very low saturation magnetization compared to Ni(Mo)/Ti have been developed, and may serve as an alternative. In this paper, we test the performance of both mirrors. At a first stage, we present four-states polarized neutron reflectivity curves of Ni(Mo) and Cu monolayers measured at the neutron reflectometer SuperADAM and perform a full polarization analysis, showing a difference between the magnetic scattering length density (mSLD) of both materials, with Cu having a lower mSLD than Ni(Mo). These results are later corroborated with the full polarization analysis of four-states polarized neutron reflectivity curves of $m=2$ Ni(Mo)/Ti and Cu/Ti supermirrors performed in the same reflectometer. In a second stage, we measured the depolarization ($D$) that a polarized neutron beam suffers after reflection from the same Ni(Mo)/Ti and Cu/Ti supermirrors by using the Opaque Test Bench setup. We find upper limits for the depolarization of $D_\text{Cu/Ti(4N5)}<\num{7.6e-5}$, $D_\text{Ni(Mo)/Ti}<\num{8.5e-5}$, and $D_\text{Cu/Ti(2N6)}<\num{6.0e-5}$ at the $1\sigma$ confidence level, where (4N5) corresponds to a Ti purity of $\SI{99.995}{\percent}$ and (2N6) to $\SI{99.6}{\percent}$. The uncertainties are statistical. These results show that all three supermirrors are suitable for being used in next generation $\beta$ decay experiments. We found no noticeable dependence of the depolarization on the $q$ value or the magnetizing field, in which the samples were placed.
  \end{abstract}


\maketitle

\section{Introduction}
Neutron mirrors are crucial elements for transporting neutrons from their origin in the neutron source to the experimental site, which may be located up to a hundred meters away. Due to the neutron optical potential of the composing materials, neutrons that are incident to the surface with an angle smaller than the critical angle $\theta_\text{crit}$ are totally reflected. By stacking multiple layers of neutron reflecting materials with different optical potentials and varying layer thickness, $\theta_\text{crit}$ can be extended beyond that of the commonly used mirror material, nickel. These systems are called supermirrors (SMs) \cite{MD77}. They are characterized by the so-called $m$-value, which is $\num{1}$ for Ni, corresponding to a critical angle of $\theta_\text{crit,Ni}[\si{\degree}]= \num{0.099}\cdot\lambda[\si{\angstrom}]$. Modern SMs can reach $m$-values up to $\num{8}$ \cite{Sch16}, extending their $\theta_\text{crit}$ to $\theta_\text{crit,SM}=\theta_\text{crit,Ni}\cdot m$. Therefore, supermirrors are used in neutron scattering and particle physics experiments to increase the transported neutron flux.\\
In the case of non-polarized neutrons, SMs are usually made of Ni and Ti, as they have a high scattering contrast \cite{Erk08}. However, due to the ferromagnetism of Ni, the reflection of a polarized neutron on a Ni/Ti mirror might flip the neutron's spin by spin-flip scattering, if there is a magnetization component out of the mirror plane \cite{Hol+22}. This makes Ni/Ti supermirrors not suitable for the transport of polarized neutron beams, as needed for the new neutron decay spectrometer PERC \cite{Dub08,Wan+19}, which is currently being set up at the research reactor FRM II in Garching, Germany. PERC aims to measure different correlation coefficients in free neutron decay up to one order of magnitude more precisely than current best values to test the Standard Model and look for physics beyond it. In PERC's \SI{8}{\metre} long decay volume, an $m=2$ SM neutron guide will be installed to keep the beam from diverging. However, in order to measure for example the beta asymmetry $A$ (asymmetry of the angular distribution of the decay electron with respect to the neutron spin) to the intended accuracy, the neutron beam is polarized before entering the spectrometer and its polarization has to stay close to constant, requiring low-depolarizing SMs with a depolarization probability of less than or equal to $\num{e-4}$ per reflection.\\
The common choice in neutron SMs for experiments with polarized neutrons is Ni(Mo)/Ti \cite{Cab86,Pad04,Kov08}, as alloying nickel with molybdenum with more than $\SI{12}{\weight\percent}$ decreases its Curie temperature to below $\SI{4}{\kelvin}$ \cite{KAN97}. Nevertheless, a small residual magnetization of Ni(Mo)/Ti SMs remains, especially in high magnetic fields. There have been some studies on spin-dependent reflectivity and depolarization of Ni(Mo)/Ti SMs \cite{Sch99,Reb14,Bon17}, however up to now, to the best of our knowledge, there have been no experiments to measure its depolarization probability for cold neutrons to the required precision.\\
To overcome the tiny, but non-negligible, magnetization of the Ni(Mo) alloy, coatings made of diamagnetic copper and paramagnetic titanium have also been investigated. Pleshanov et al. \cite{Ple94} first reported on the roughness and interdiffusion of samples with $20$ bilayers of Cu and Ti thermally evaporated on glass substrates. The depolarization of the first Cu/Ti SMs \cite{Reb14}, which reached $m=1.2$, was investigated with the Opaque Test Bench (OTB) setup \cite{Kla12,Kla13}, described in detail later. The ratio between the direct beam polarization and the polarization measured after reflection from the Cu/Ti sample indicated no depolarization down to the order of $\num{e-4}$ for $\SI{5.3}{\angstrom}$ neutrons. Hollering at al. \cite{Hol+22} produced an $m=2$ Cu/Ti SM with non-polarized reflectivity above $\SI{90}{\percent}$ with RF magnetron sputtering (Cu) and pulsed DC magnetron sputtering (Ti). While this mirror could not be tested with polarized neutrons, newer Cu/Ti SMs produced by the Neutron Optics Group of the FRM II show a successful polarized neutron reflectivity (PNR) of $\SI{90}{\percent}$ for $\SI{4.67}{\angstrom}$ polarized neutrons \cite{Gom24}.\\
In the first part of this paper we report on 4-states PNR measurements performed at the neutron reflectometer SuperADAM \cite{Anton13,SADAM} at the Institut Laue Langevin (ILL) in Grenoble, France, with the aim to extract information about the magnetic moment of Cu and Ni(Mo) monolayers and Cu/Ti and Ni(Mo)/Ti SMs. In the second part of the paper, the depolarization that a polarized neutron beam suffers after reflection from a Ni(Mo)/Ti SM and two Cu/Ti SMs produced using titanium with different degrees of purity is experimentally determined with the Opaque Test Bench setup \cite{Kla12,Kla16} at the PF1B beamline of the ILL \cite{Pf1b,Abe06}. Data is available at \cite{Dat23}.

\section{Samples}
All samples investigated in this paper were produced in the same conditions as the ones of \cite{Gom24}. The $\SI{1500}{\angstrom}$ Cu and $\SI{800}{\angstrom}$ Ni(Mo) (alloy $\num{83}/\SI{17}{\weight\percent}$) monolayers and the Cu/Ti and Ni(Mo)/Ti $m=2$ neutron supermirrors were produced using an in-house DC magnetron sputtering system. For all depositions reported here high purity $\SI{99.99}{\percent}$ Cu, Ni(Mo) and Ni, and low purity $\SI{99.6}{\percent}$ (2N6 from now on) Ti sputtering targets manufactured by Metallic Flex GmbH as well as a high purity $\SI{99.995}{\percent}$ (4N5 from now on) Ti sputtering target manufactured by Evochem Advanced Materials GmbH were used. Test samples were deposited by DC magnetron sputtering at a constant power of $\SI{900}{\watt}$ (Cu), $\SI{800}{\watt}$ (Ni and Ni(Mo)) and $\SI{750}{\watt}$ (Ti). All samples were sputtered on Borofloat\textsuperscript{\textregistered} 33 glass manufactured by Schott AG (Germany), with dimensions $\qtyproduct{50x50x1}{\milli\metre}$ (monolayers) and $\qtyproduct{50x50x11}{\milli\metre}$ (SMs). The layer sequence for the different SMs with gradually varying layer thicknesses was calculated using the algorithm of Masalovich \cite{Mas}.\\
The samples were previously characterized with different experimental techniques, investigating the influence of adding synthetic air during the sputtering process on the evolution of the roughness and the formation of different crystal structures in the single layers. The magnetic properties of the Cu/Ti and Ni(Mo)/Ti SMs were measured by Superconducting Quantum Interference Device (SQUID) magnetometry. Copper was found to have a magnetic moment of only a twentieth of the commonly used Ni(Mo) alloy. Depolarization simulations using the data obtained by SQUID magnetometry suggested that both the Ni(Mo)/Ti and Cu/Ti SMs would depolarize below $\num{e-4}$, with the Cu/Ti SM performing better by a factor of $100$, when no instrumental depolarization effects are included within the simulation. Details can be found in \cite{Gom24}.

\section{SuperADAM}
\subsection{Setup}
Four-states PNR curves were measured at the SuperADAM instrument at the Institut Laue-Langevin, Grenoble, France, \cite{Anton13,SADAM} and a full polarization analysis was performed. The instrument uses a solid state polarizer/wavelength filter providing a highly polarized (up to $\SI{99.7}{\percent}$) monochromatic neutron beam with wavelength $\SI{5.183}{\angstrom}$, monochromatization $\Delta\lambda/\lambda=\SI{0.5}{\percent}$ (FWHM) and angular divergence $\Delta\alpha = \SI{0.33}{\milli\radian}$ (FWHM).  For polarization analysis, we chose the multi-mirror analyzer option with a polarization analysis efficiency of $\SI{99.3}{\percent}$ for the monolayer measurements. For the SM samples, we chose the $^{3}$He analyzer option ($p_\text{He} = \SI{1.35}{\bar}$ and $P_\text{He}=\SI{78.2}{\percent}$). The flipping of the incident neutron polarization is performed using a $\SI{180}{\degree}$ adiabatic gradient radio frequency flipper with an efficiency of up to $\SI{99.97}{\percent}$. The samples were placed in a vertical $\SI{0.8}{\tesla}$ magnetic field parallel to the field of the polarizer.

\subsection{Results}
\subsubsection{Monolayers}
\begin{figure*}[t]
    \centering
		\includegraphics[width=0.8\textwidth]{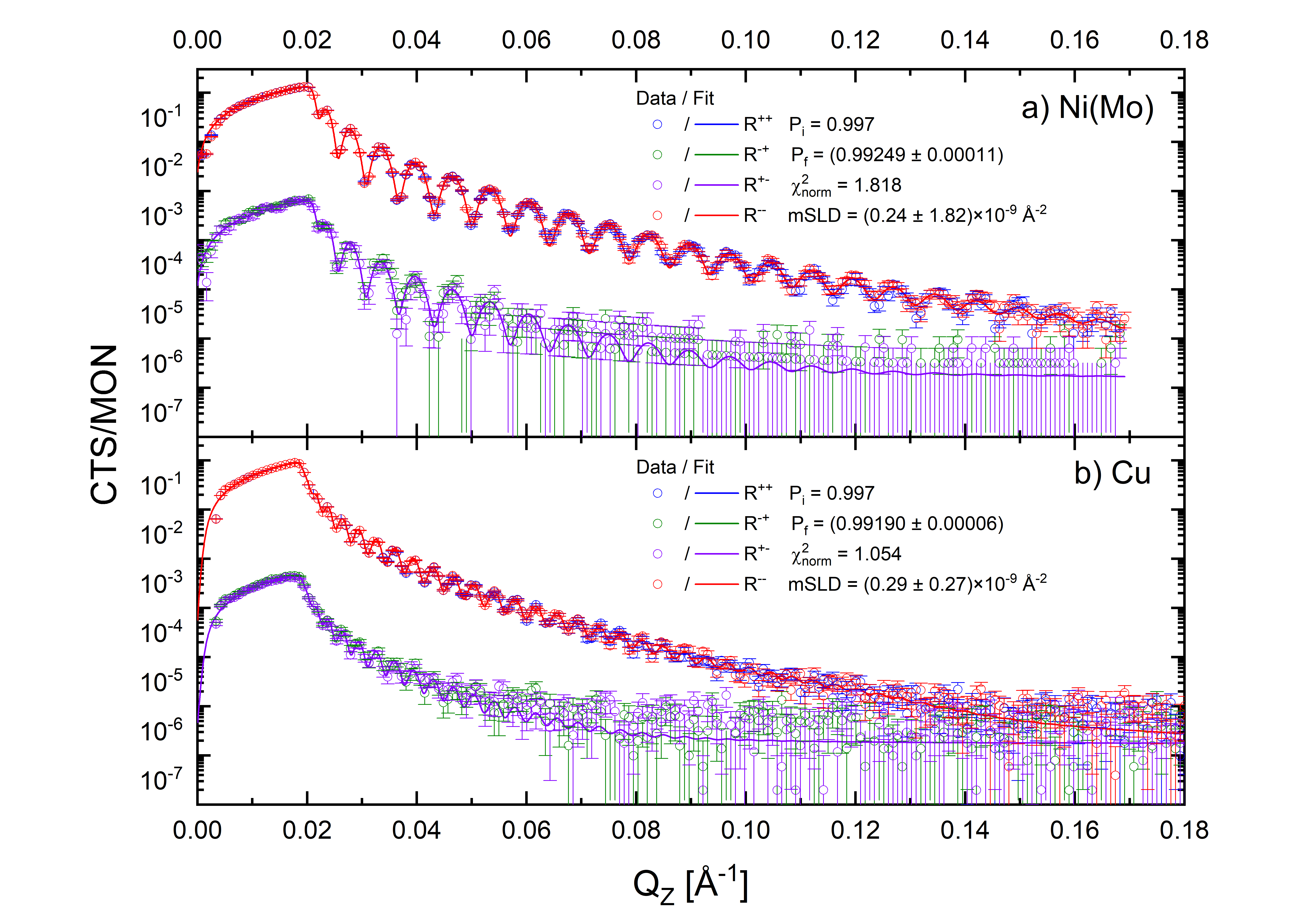}
  \caption{Experimental reflectivity curves for R$^{++}$ (blue), R$^{--}$ (red), R$^{+-}$ (purple) and R$^{-+}$ (green) spin states recorded for a $\SI{800}{\angstrom}$ thin Ni(Mo) film (top) and a $\SI{1500}{\angstrom}$ thin Cu film (bottom) as measured in SuperADAM. Solid lines are the best fits to the experimental data. The incident polarization was set constant at $P_{i} = 0.997$, other parameters, such as the analyzing efficiency, thickness, SLD and mSLD of the monolayer samples, and the SLD of the substrate were allowed to vary.}
  \label{SAmono}
\end{figure*}

Figure \ref{SAmono} shows 4-states PNR curves with full polarization analysis from the $\SI{800}{\angstrom}$ Ni(Mo) (top) and the $\SI{1500}{\angstrom}$ Cu monolayer samples (bottom), performed with the multi-mirror analyzer. The figure includes the result of the least square fits performed with the original software BoToFit \cite{Anton13}, developed at ILL, which permits simultaneous fitting of up to four - two non-spin-flip (NSF) and two spin-flip (SF) - reflectivity curves in each iteration cycle. The algorithm of PNR calculation is based on the super-iterative routine \cite{Super}. The least square routine is applied to the theoretical curves that were convoluted with the instrumental resolution function and corrected for polarization efficiencies.
During the fitting process, the incident polarization was set as a constant parameter $P_{i} = 0.997$, while other parameters such as the analyzing efficiency ($P_{f}$), thickness, Scattering Length Density (SLD) and Magnetic Scattering Length Density (mSLD) of the monolayer samples, and the SLD of the substrate were allowed to vary. The fitting yields an analyzing efficiency of $P_f = (0.99190 \pm 0.00006)$ for the Cu monolayer sample and $P_f = (0.99249 \pm 0.00011)$ for the Ni(Mo) monolayer sample, both in reasonable agreement with the technical specification of the instrument $P_f = 0.993$.\\
In the case of the $\SI{800}{\angstrom}$ Ni(Mo) sample (figure \ref{SAmono}, top), the fitting yields a SLD of $(8.675 \pm 0.003)\times 10^{-6}\si{\per\square\angstrom}$, which compares well to the value of $\SI{8.26e-6}{\per\square\angstrom}$, as given by \cite{Kov08} for this Ni(Mo) alloy. The mSLD was found to be $(0.24 \pm 1.82)\times 10^{-9}\si{\per\square\angstrom}$, which converts into a magnetization of $(0.08 \pm 0.64)$ emu/cm$^{3}$, in good agreement with the value obtained by SQUID magnetometry of the Ni(Mo) sputtering target reported by \cite{Gom24}. However, the large uncertainty indicates a detection limit in the mSLD for the PNR technique. The overall normalized ${\chi}^2$ of the fitting was found to be 1.818. The substrate showed a SLD of $(3.31 \pm 0.04)\times 10^{-6}\si{\per\square\angstrom}$, which compares very well to the value of $\SI{3.40e-6}{\per\square\angstrom}$, given by \cite{Wel} for SiO$_\text{2}$, which makes $\SI{81}{\percent}$ of the chemical composition of Borofloat\textsuperscript{\textregistered} 33. For the $\SI{1500}{\angstrom}$ Cu monolayer sample shown in figure \ref{SAmono} (bottom), the fitting yields a SLD of $(6.455 \pm 0.002)\times 10^{-6}\si{\per\square\angstrom}$, which compares very well to the value of $\SI{6.55e-6}{\per\square\angstrom}$, given by \cite{Wel} for Cu. The mSLD was found to be $(0.29 \pm 0.27)\times 10^{-9}\si{\per\square\angstrom}$, which converts into a magnetization of $(0.10 \pm 0.09)$ emu/cm$^{3}$. This is larger than the value obtained by SQUID magnetometry in the Cu sputtering target reported by \cite{Gom24}, but the deviation is very small compared to the large uncertainty of the value resulting from the PNR fit. The large uncertainty of the mSLD obtained for the Cu sample again indicates a detection limit in the mSLD for the PNR technique. The overall normalized ${\chi}^2$ of the fitting was found to be 1.054. The substrate showed a SLD of $(3.38 \pm 0.03)\times 10^{-6}\si{\per\square\angstrom}$, which compares very well to the value of $\SI{3.40e-6}{\per\square\angstrom}$, given by \cite{Wel} for SiO$_\text{2}$.\\

\subsubsection{Supermirrors}
\begin{figure*}[t]
    \centering
		\includegraphics[width=0.8\textwidth]{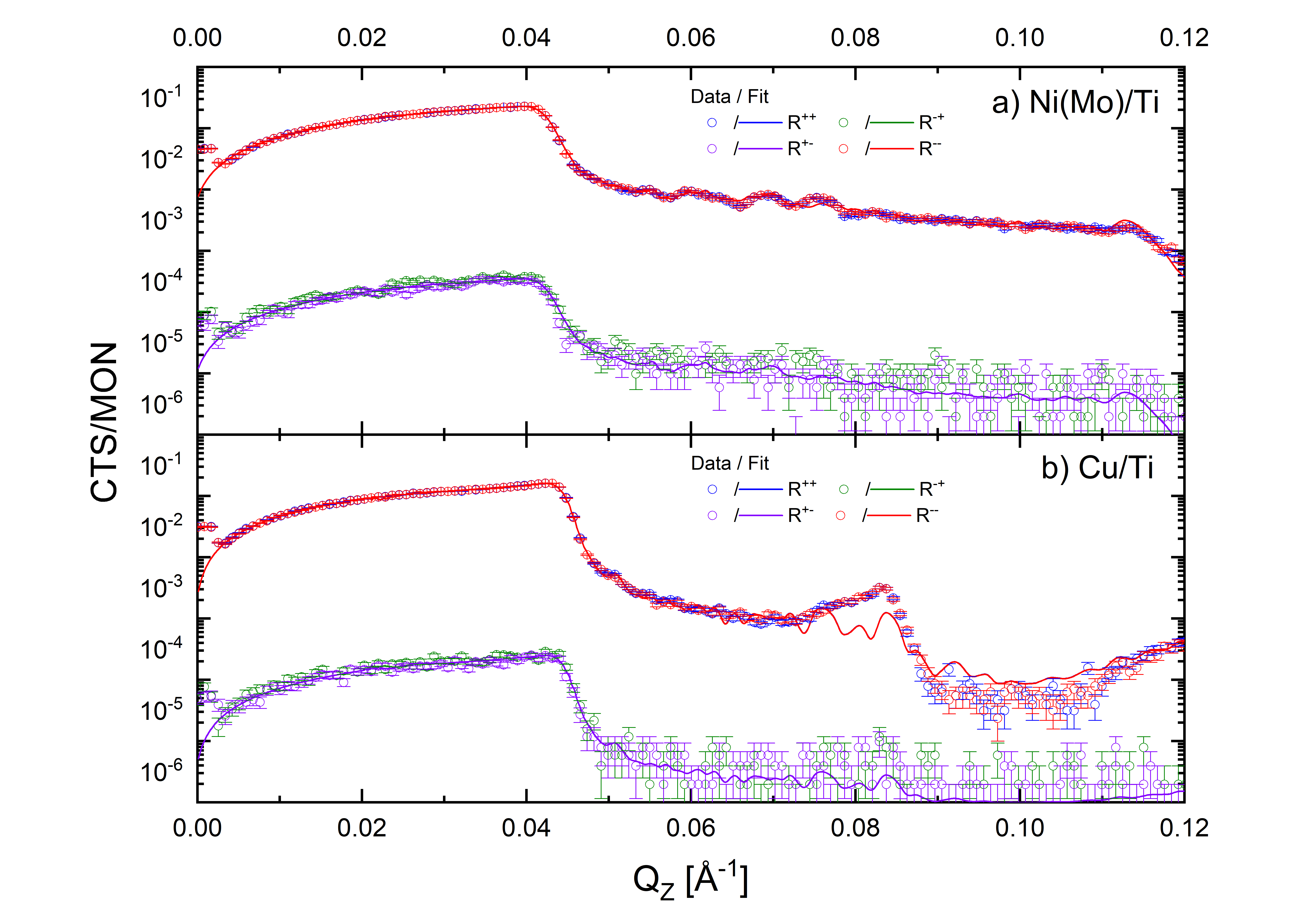}
  \caption{Experimental reflectivity curves for R$^{++}$ (blue), R$^{--}$ (red), R$^{+-}$ (purple) and R$^{-+}$ (green) spin states recorded for $m=2$ Ni(Mo)/Ti SM (top) and Cu/Ti $m=2$ SM (bottom), as measured in SuperADAM. Solid lines are the best fits to the experimental data. The fit was achieved in two stages due to the large number of layers.}
  \label{SM}
\end{figure*}

Figure \ref{SM} shows 4-states PNR curves with full polarization analysis for the Ni(Mo)/Ti (top) and the Cu/Ti (bottom) $m=2$ SM coated samples. The figure also includes the result of the least square fits performed with a custom fitting routine designed specifically for SM structures. Due to the rather large number of layers in the SM, the routine is performed in two stages with an initial fit with a small number of parameters such as the SLD of the different materials, roughness and simple linear gradient variation of the stack. Once the initial fitting stage converges to a minimum, a full stack layer model with more than 600 parameters is fitted. The same algorithm for full polarization analysis and instrument resolution convolution as for the monolayers is applied.\\

\begin{figure}[h]
    \centering
		\includegraphics[width=0.48\textwidth]{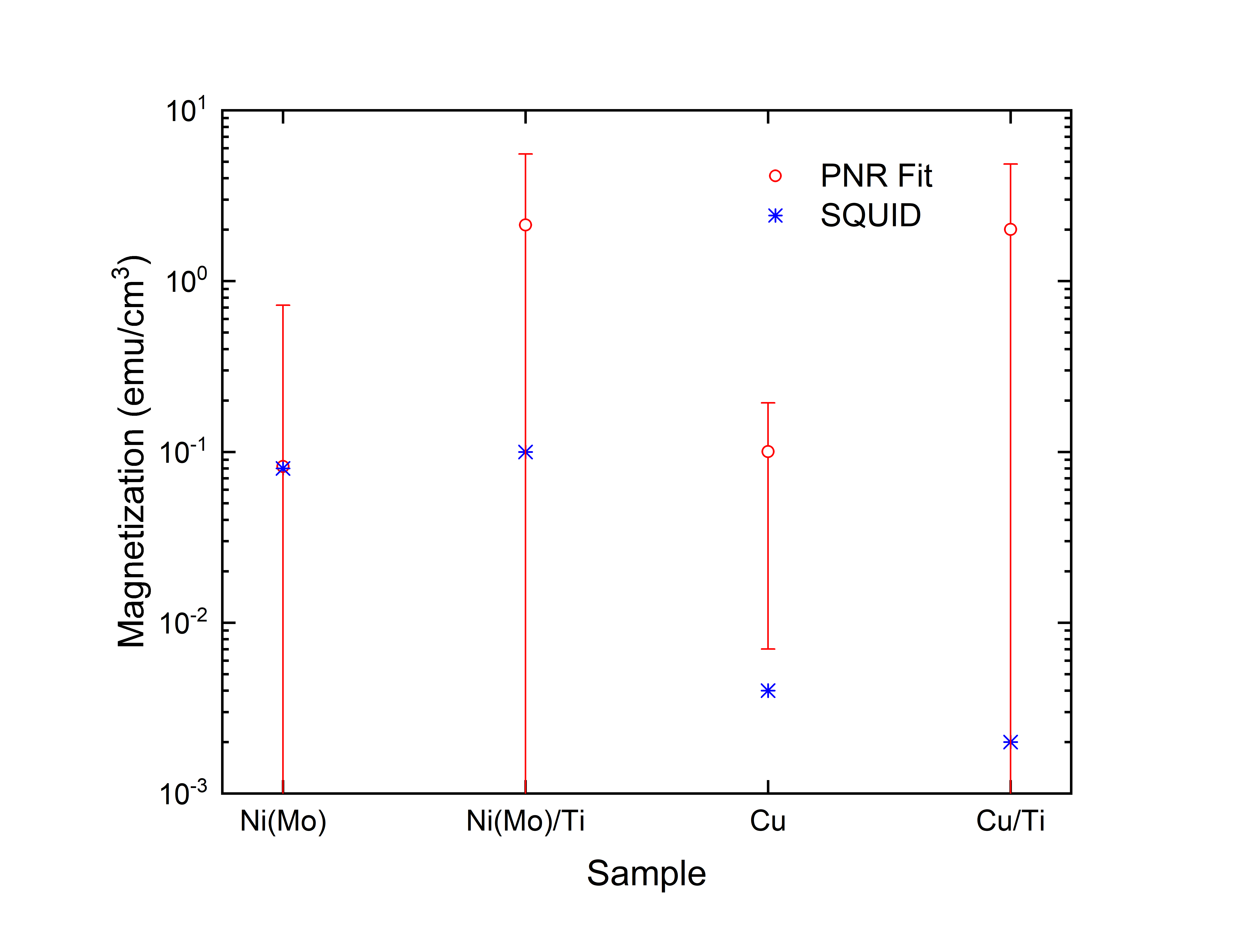}
  \caption{Experimentally measured magnetization of different coatings. The red dots represent the average magnetization of the coating obtained from the fit of the measured set of polarized reflectivity curves. The blue stars represent the magnetization obtained from the SQUID probe \cite{Gom24}, where the errors of $\approx\SI{1}{\percent}$ are too small to be visible. Cu and Ni(Mo) here are designating the monolayer coatings while the Cu/Ti and Ni(Mo)/Ti are the thick multilayer stack of the corresponding supermirrors. Except for Cu, the values agree within the errors.}
  \label{SASum}
\end{figure}

Due to the complexity of the SM structure and the low number of experimental points, the fitting quality is significantly worse than for the monolayer samples above. The large number of layers introduces significant lateral correlations leading to significant off-specular scattering for both magnetic and nuclear scattering length profiles. Therefore, even a plausible fit of the structure produces an imperfect model and parameter errors that are significantly higher than of the simple models above. We obtain a magnetization of $(2.1 \pm 3.4)$ emu/cm$^{3}$ for Ni(Mo)/Ti and $(2.0 \pm 2.8)$ emu/cm$^{3}$ for Cu/Ti. The thicknesses of the individual layers of both Cu/Ti and Ni(Mo)/Ti found by the fitting of the PNR SMs samples deviate from the theoretical thickness by less than $\SI{2}{\percent}$. The mSLD has been considered constant for each material throughout the stack.\\
Figure \ref{SASum} depicts the final magnetization results (in emu/cm$^{3}$) after converting the mSLD into units comparable with SQUID magnetometry. The values reported by \cite{Gom24} are presented for reference. The results show that at such low level of magnetic signal the PNR struggles to produce significant results, as can be seen by rather large statistical uncertainty of the final values.

\section{The Opaque Test Bench}
\subsection{Setup}
\begin{figure}[h]
    \centering
		\includegraphics[width=0.48\textwidth]{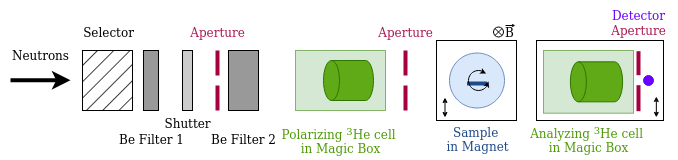}
  \caption{Schematic of the OTB setup. The magnet and sample are placed on a stage that can be moved in and out of the beam, and rotated around the central axis. The magnetic field along the neutron beam between polarizer and analyzer was perpendicular to the plane of the drawing. The analyzing cell together with the detector aperture and detector can be moved to be able to measure the direct and the reflected beams. The Beryllium filter 2 was only present during the final measurements.}
  \label{otbsetup}
\end{figure}

The Opaque Test Bench (OTB) \cite{Kla12,Kla13,Kla16} is currently the only setup to test the depolarization of neutron mirrors down to the level of $\mathcal{O}(\num{e-4})$ due to the high polarizing power of the two cells filled with polarized $^{3}$He used as polarizer (P) and analyzer (A). A schematic of the setup is shown in fig. \ref{otbsetup}. The cells are placed inside so-called magic boxes \cite{Pet06}, which provide an homogeneous magnetic field for the polarized $^3$He and shield from external magnetic fields using $\mu$ metal. Additionally, an adiabatic fast passage spin flipper is installed in each box, to flip the spin of the $^3$He as needed with negligible losses of the degree of $^3$He polarization $P_\text{He}$ on the order of $\num{4e-5}$ per flip, which translates to a reduction of the polarizing or analyzing power for $\SI{5}{\angstrom}$ neutrons of $\num{2e-9}$ per flip. The expected product of polarizing power $P$ and analyzing power $A$ from the two identical $^{3}$He cells can be calculated as

\begin{equation}
    AP = \tanh{\left(P_\text{He}O\right)}\cdot \tanh{\left(P_\text{He}O\right)}= \tanh^2{\left(P_\text{He}O\right)},
\end{equation}

where $O=O(\lambda)$ is the so-called opacity of the cell \cite{Cou88,Kla12}. In principle, $P_\text{He}$ and $O$ may be different for polarizer and analyzer, but in our case we assume both are equal since the cells were similar and were produced close to each other in time.\\
In the experiment presented in this paper, a specific wavelength range from the cold neutron beam coming from the reactor was selected by using a Dornier velocity selector \cite{Fri89}, getting a wavelength resolution of $\delta\lambda/\lambda\approx\SI{10}{\percent}$ (FWHM). Then two apertures were used to define the beam cross section and to limit the beam divergence. A third aperture directly in front of the detector was used to separate direct and reflected beam. The count rate $N$ was measured with a $^3$He detector placed behind the analyzer for two cases: the $^3$He polarization in the analyzing cell is parallel (from now on white configuration) or anti-parallel (from now on black configuration) to that of the polarizing cell. There were two possible spin combinations for both cases: $\uparrow\uparrow$ and $\downarrow\downarrow$ for white, and $\downarrow\uparrow$ and $\uparrow\downarrow$ for black. The product $AP$ is then given by

\begin{equation}
    AP=\dfrac{\left(N_\text{w}-N^\text{bg}_\text{w}\right)-\left(N_\text{b}-N_\text{b}^\text{bg}\right)}{\left(N_\text{w}-N^\text{bg}_\text{w}\right)+\left(N_\text{b}-N_\text{b}^\text{bg}\right)},
    \label{AP}
\end{equation}

where $N_\text{w}$ and $N_\text{b}$ denote the uncorrected count rate in the white ($N_\text{w}$ = $N_{\uparrow\uparrow} + N_{\downarrow\downarrow}$) and black ($N_\text{b}$ = $N_{\downarrow\uparrow} + N_{\uparrow\downarrow}$) configurations, and the superscript $\text{bg}$ refers to the same parameters for the background measurements, performed with the shutter in figure \ref{otbsetup} closed.\\
To determine the degree of depolarization after reflecting off a sample, a supermirror is placed in between the analyzer and the polarizer, so that the neutron beam is reflected at an angle below $\theta_\text{crit,SM}$. The sample is placed inside an electromagnet that generates a magnetizing field parallel to the mirror surface and perpendicular to the beam direction. Analogous to $AP$, the product $ASP$ can be determined with Eq. \eqref{AP} by measuring the count rate for the two cell configurations.\\
The associated error of $AP$ ($A S P$) comes from error propagation, assuming Poisson errors for the number of counts (${n}$) used to determine the count rates (${N}$) in the different cell configurations. Including a confidence limits approach \cite{gehrels} for the cases with small numbers of counts (like the black configuration or the background measurements) does not introduce any noticeable difference as compared to the general case where the errors of the number of counts $n$ are treated as $\sqrt{n}$. With $AP$ and $ASP$, the depolarization of the neutron beam induced by the SM sample is determined as \cite{Reb14}

\begin{equation}
    D=1-\left(ASP/AP\right),
    \label{depol}
\end{equation}

with the associated error again calculated from error propagation.

\subsection{Measurement Procedure}
The sample was held in place in between the pole shoes of an electromagnet by a holder made out of aluminum, so that the magnetic field is undisturbed. The magnetic field was measured in the center of the magnet at five different currents and interpolated linearly between those points. The magnet was able to generate a magnetic field up to $\SI{0.66}{\tesla}$, which is of the same order as the magnetic field inside the decay volume of PERC (between $\SI{0.5}{\tesla}$ and $\SI{1.5}{\tesla}$). Guide fields with a minimum magnetic field of $\SI{10}{\gauss}=\SI{1}{\milli\tesla}$ were installed where needed to assure spin transport after passage through the polarizer.\\
For each measurement of the product $AP$ ($ASP$) the same protocols, as listed in table \ref{otbtimes}, were followed. All four combinations were measured twice in each protocol in the order given in table \ref{otbtimes}, to account for possible drifts, in particular the relaxation time of $P_\text{He}$. Due to the much lower count rate in the black configurations, these were measured $10$ times longer. 
\begin{table}[h]
  \caption{Order and measurement times of the different configurations in the short and long protocols.}
  \begin{tabular}{|l|ll|ll|}
  \hline
  \multicolumn{1}{|c|}{\multirow{2}{*}{Run}} & \multicolumn{2}{c|}{Direction of $^3$He pol.} & \multicolumn{2}{c|}{Duration [s]} \\ \cline{2-5} 
\multicolumn{1}{|c|}{} & \multicolumn{1}{c|}{Polarizer} & Analyzer & \multicolumn{1}{c|}{\begin{tabular}[c]{@{}c@{}}Short\\ protocol\end{tabular}} & \multicolumn{1}{c|}{\begin{tabular}[c]{@{}c@{}}Long\\ protocol\end{tabular}} \\ \hline
0 & \multicolumn{1}{c|}{$\uparrow$} & \multicolumn{1}{c|}{$\uparrow$} & \multicolumn{1}{l|}{20} & 80 \\ \hline
1 & \multicolumn{1}{c|}{$\downarrow$} & \multicolumn{1}{c|}{$\uparrow$} & \multicolumn{1}{l|}{200} & 800 \\ \hline
2 & \multicolumn{1}{c|}{$\downarrow$} & \multicolumn{1}{c|}{$\downarrow$} & \multicolumn{1}{l|}{20} & 80 \\ \hline
3 & \multicolumn{1}{c|}{$\uparrow$} &\multicolumn{1}{c|}{ $\downarrow$} & \multicolumn{1}{l|}{200} & 800 \\ \hline
4 & \multicolumn{1}{c|}{$\uparrow$} &\multicolumn{1}{c|}{ $\downarrow$} & \multicolumn{1}{l|}{200} & 800 \\ \hline
5 & \multicolumn{1}{c|}{$\downarrow$} & \multicolumn{1}{c|}{$\downarrow$} & \multicolumn{1}{l|}{20} & 80 \\ \hline
6 & \multicolumn{1}{c|}{$\downarrow$} & \multicolumn{1}{c|}{$\uparrow$} & \multicolumn{1}{l|}{200} & 800 \\ \hline
7 & \multicolumn{1}{c|}{$\uparrow$} & \multicolumn{1}{c|}{$\uparrow$} & \multicolumn{1}{l|}{20} & 80 \\ \hline
  \end{tabular}
  \label{otbtimes}
  \end{table}
Regular background measurements were performed for either $\SI{20}{\second}$ in $\uparrow\uparrow$ and $\SI{200}{\second}$ in $\downarrow\uparrow$ (short protocol) or $\SI{80}{\second}$ in $\uparrow\uparrow$ and $\SI{800}{\second}$ in $\downarrow\uparrow$ (long protocol), during which the beam shutter installed after the velocity selector was closed. These measurements were taken for the OTB setup to either measure the products $AP$ and $ASP$, and lead to the correction in Eq. \ref{AP} by the terms $N_i^{\rm bg}$.\\
The $^3$He cells used in this study were prepared by the Neutron Optics Group of the ILL with their TYREX-2 filling station \cite{TYR2}. The cells had a length of $\SI{15}{\centi\metre}$ each and the $^3$He had a $P_\text{He}$ between $\SI{78.7}{\percent}$ and $\SI{79.7}{\percent}$ after filling. As the $^3$He depolarizes over time, the cells were exchanged by new ones at least once a day. The results shown in this work were taken within the first $\SI{12}{\hour}$ after the installation of the $^3$He cells. By checking the direct beam count rate decrease over time in the $\uparrow\uparrow$ configuration, we confirm that the relaxation time of the $^3$He polarization of the cells within the magic boxes is well above $\SI{200}{\hour}$. This results in a loss of $AP$ ($ASP$) on the order of a few $\num{e-6}$ in the course of our measurements, which is low enough compared to the statistical error not to require further time correction. Note, however, that the reduced transmission reduces the statistics per time and the signal/background ratio. The neutron detector used, a $^{3}$He tube of $\SI{10}{\milli\metre}$ diameter, had an efficiency of $\approx{\SI{53}{\percent}}$ at $\lambda=\SI{1.8}{\angstrom}$ and was shielded from ambient background by borated rubber and borated polyethylene blocks.\\
Wavelengths between $\SI{4.5}{\angstrom}$ and $\SI{6}{\angstrom}$ were selected for the experiment, staying close to $\SI{5}{\angstrom}$, which is planned to be selected for measurements with PERC. The neutron wavelength spectra after the selector were measured without cells by time-of-flight measurements using a disk chopper and corresponded to the expectations for a selector axis aligned parallel to the beam axis (relative deviation of measured central wavelength from expectation $\mathcal{O}(\num{e-3})$).

\subsection{Results}
\begin{figure}[h]
  \centering
  \includegraphics[width=0.48\textwidth]{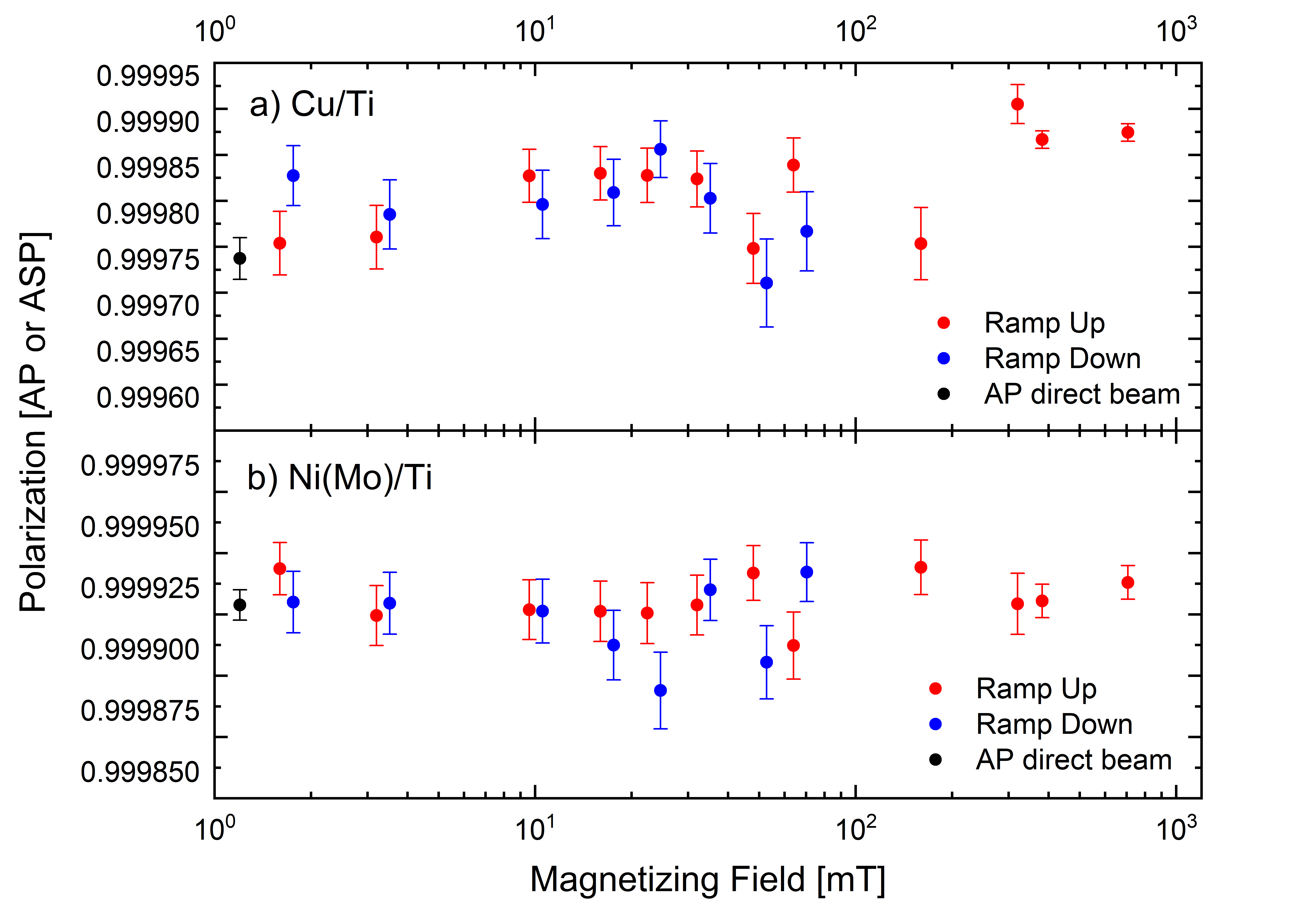}
\caption{$AP$ of the direct beam (black dots) and $ASP$ of the neutron beam after reflection from the Cu/Ti(4N5) sample (top) and the Ni(Mo)/Ti sample (bottom) for different magnetic fields with $\lambda=\SI{5}{\angstrom}$ and $\theta = \SI{0.73}{\degree}$ while ramping up (red dots) and down (blue dots). The values when ramping down are plotted slightly shifted in $x$ for better visibility. Neither sample shows any noticeable dependence on the magnetic field strength, indicating that both samples are already fully magnetized at the minimum field. The data points of the two highest fields come from one long measurement each, all others are the result of one short measurement per point.}
\label{bfield}
\end{figure}

Polarization measurements of the neutron beam after reflection in the Cu/Ti(4N5) sample were taken at three different nominal cell pressures ($\SI{1.4}{\bar}$, $\SI{1.6}{\bar}$ and $\SI{1.8}{\bar}$) and at least at two different selector speeds. The chosen selector speeds correspond to neutron wavelengths $\SI{4.5}{\angstrom}$, $\SI{5}{\angstrom}$, and $\SI{5.5}{\angstrom}$. The incident angle of the incoming neutron beam was set to $\theta=\SI{0.73}{\degree}$ to be within the range of total reflection for $m=2.0$ SMs for $\SI{5}{\angstrom}$ neutrons. From the collected data it was found that a pressure of $\SI{1.6}{\bar}$ and a selected wavelength of $\SI{5}{\angstrom}$ are ideal for the depolarization measurements, showing both a very high polarization and a reasonable transmission of the polarized neutron beam. Therefore, all results shown in the following are taken with cells of that pressure and a selector speed corresponding to $\SI{5}{\angstrom}$.\\
Polarization measurements with both the Cu/Ti(4N5) and the Ni(Mo)/Ti sample were taken for different magnetizing fields at an incident angle of $\SI{0.73}{\degree}$. The field was gradually increased from $\SI{1.6}{\milli\tesla}$ to $\SI{660}{\milli\tesla}$ and decreased again. The results, depicted in figure \ref{bfield}, show that the polarization of the neutron beam after reflection off either sample is independent of the magnetizing field, indicating that both samples are already fully magnetized at the minimum field. 

\begin{figure}[h]
    \centering
		\includegraphics[width=0.48\textwidth]{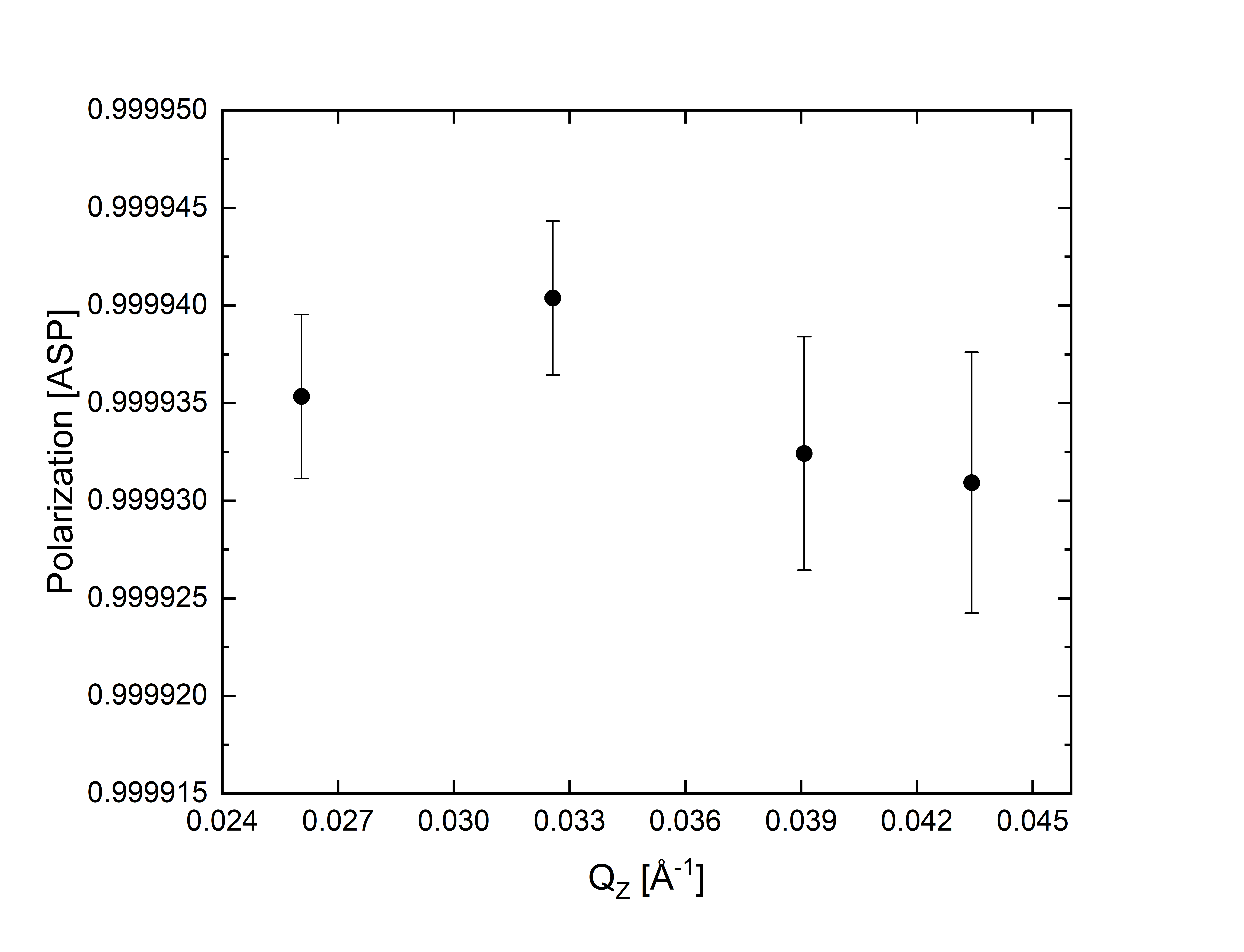}
  \caption{Polarization of beam reflected by the Cu/Ti(4N5) sample at different angles, corresponding to different momentum transfer $q$. All the values agree within the error bars, showing no sign of a dependence on the $q$ value. The wavelength was selected to be $\lambda = \SI{5}{\angstrom}$ and each point is the result of at least three short measurements or one long one.}
  \label{angle}
\end{figure}

To measure the depolarization dependence of the Cu/Ti(4N5) sample on the momentum transfer $q = \frac{4\pi}{\lambda}\sin\theta$, the polarization $ASP$ of the neutron beam after reflection was compared for four different $\theta$ angles ($\SI{0.58}{\degree}$, $\SI{0.73}{\degree}$, $\SI{0.88}{\degree}$ and $\SI{0.97}{\degree}$), all within the range of high reflectivity for $\SI{5}{\angstrom}$ neutrons, as $\theta_{\text{crit},\SI{5}{\angstrom}}=\SI{0.99}{\degree}$ for an $m=2$ SM. Figure \ref{angle} shows no dependence of the polarization of the reflected beam while changing its momentum transfer $q$, as all values agree within the error bars.

\begin{figure}[h]
    \centering
		\includegraphics[width=0.48\textwidth]{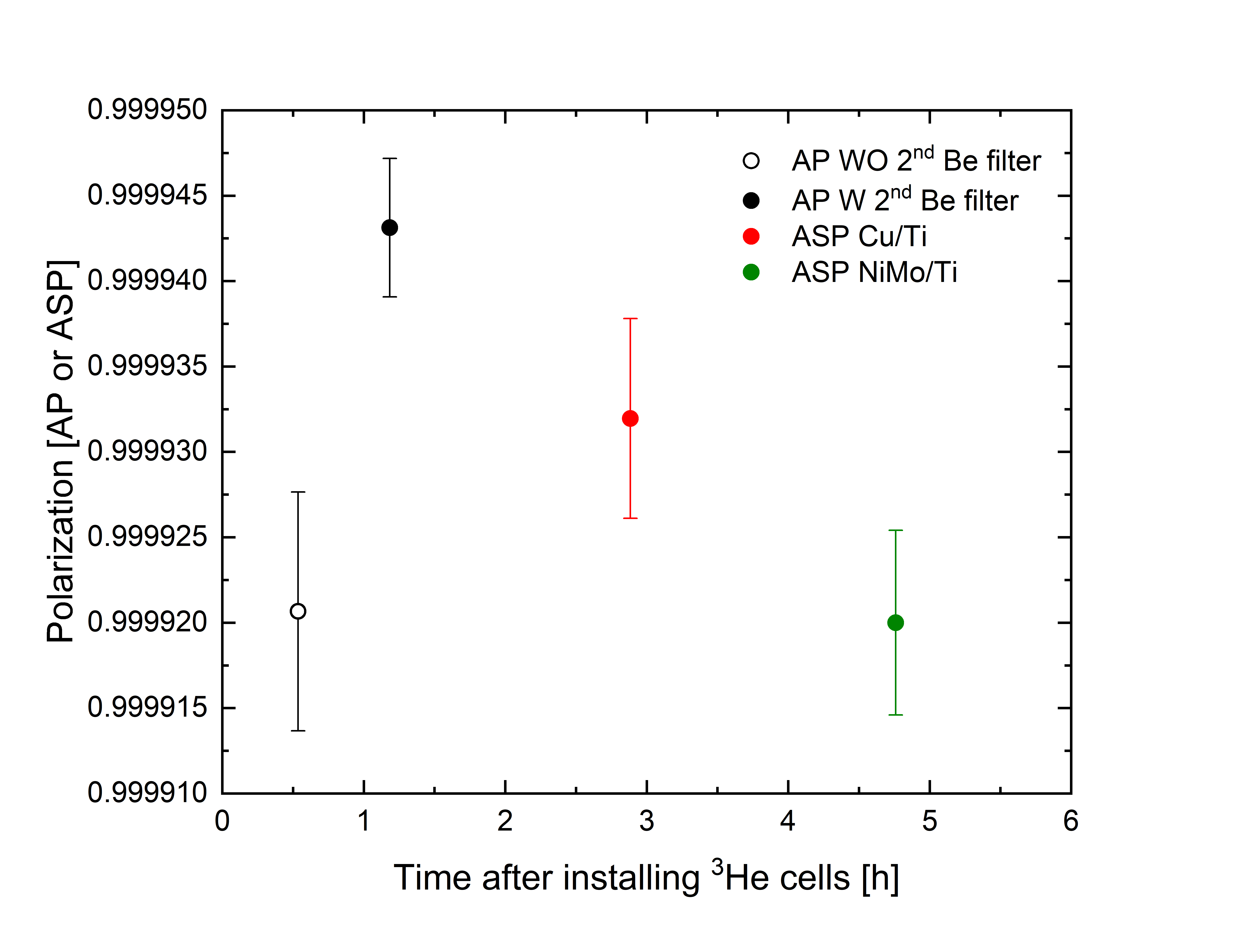}
  \caption{$AP$ at $\lambda = \SI{5}{\angstrom}$ with (black dot) and without (open black dot)the Beryllium Filter 2 over time after the installation of the $^3$He cells. Also shown are $ASP$ of Cu/Ti(4N5) (red dot) and Ni(Mo)/Ti (green dot) at $\theta = \SI{0.73}{\degree}$. The installation of the filter improves $AP$ and increases it above the $ASP$ values of both samples.}
  \label{polwBe}
\end{figure}

During the experiment, it was found that the direct beam had a significant short wavelength neutron contribution around $\SI{1}{\angstrom}$ after the velocity selector. The short wavelength neutrons are not reflected by the SM samples, resulting in $AP<ASP$, so only relative statements in the reflected beam can be made for the data in which the short wavelength neutron contamination of the direct beam is present. A $\SI{4}{\centi\metre}$ thick Beryllium filter at room temperature similar to the one used in \cite{Kla13} was found to be insufficient in suppressing the short wavelength neutron contamination. In the following, we present measurements using a second $\SI{15}{\centi\meter}$ thick Beryllium filter cooled to 77K with a Peltier-cooled cryostat, which was installed in front of the polarizing cell (Be Filter 2 in figure \ref{otbsetup}). The effect of this second Beryllium filter on the $AP$ of the direct beam is discussed in \ref{appfnc}.\\
With the Beryllium Filter 2 installed, a decrease of about $\approx\SI{17.4}{\percent}$ in count rate in the OTB setup was observed. The polarization of the direct beam $AP$ considerably improved at $\lambda=\SI{5}{\angstrom}$. Figure \ref{polwBe} shows the measured $AP$ with (black dot) and without (open black dot) the Beryllium Filter 2 installed over the time after the installation of the $^3$He cells. With the Beryllium Filter 2 installed, the polarization of the direct beam ($AP$) is now higher than the polarization of the beam after reflection ($ASP$) on the Cu/Ti(4N5) (red dot) and the Ni(Mo)/Ti (green dot) SM samples, as measured with a reflection angle $\theta=\SI{0.73}{\degree}$.

It is now possible to calculate the depolarization $D$ of the beam after reflection off either sample with Eq. \ref{depol}, being $D_\text{Cu/Ti(4N5)}=(1.1 \pm 0.7)\times 10^{-5}$ and $D_\text{Ni(Mo)/Ti}=(2.0 \pm 0.7)\times 10^{-5}$. Although no data of the Cu/Ti(2N6) sample were taken in the OTB set-up after the installation of the Beryllium Filter 2, it is possible to compare $ASP$ of Cu/Ti(2N6) and Cu/Ti(4N5) from measurements without the filter with respect to the corresponding $AP$. By doing so, a $D_\text{Cu/Ti(2N6)}=(0.8 \pm 1.4)\times 10^{-5}$ is found, indicating that also Cu/Ti(2N6) is suitable for the use in PERC. The uncertainties on these values are only statistical.\\
The theoretical $AP$ of the direct beam can be calculated from convoluting the time of flight spectrum with the Beryllium filter 2, uncorrected by background, with the transmission and polarization by two identical $^3$He cells ($p=\SI{1.586}{\bar}$, $P_{\text{He}}=\SI{79}{\percent}$, $l=\SI{15}{\centi\metre}$) and get $AP_\textrm{theo}=0.999975(19)$ at $t=0$. Assuming that the single counts in the $\SI{0}{\angstrom}$ to $\SI{4}{\angstrom}$ region are background of the time-of-flight measurement due to, e.g., minor leakage through the chopper disk, which would not be relevant for the measurements with the $^3$He cells, $AP_\textrm{theo}$ improves to $0.999994$ with an uncertainty of $\mathcal{O}(\num{e-8})$. The difference between either $AP_\textrm{theo}$ and the measured $AP$ could be explained by unknown background from scattered neutrons in the experiment area, which are not measured when the shutter is closed, due to unknown depolarization of the $^3$He during the installation of the cells or due to a lower $^3$He pressure in the cells.\\
Given the available data, an estimate of the systematic uncertainty cannot be done. However, an upper limit for the depolarization can be set when assuming a perfectly polarized instrument, i.e., $AP =1$, which disregards all systematic uncertainties in the direct beam. This yields $D_\text{Cu/Ti(4N5)}<\num{7.6e-5}$, $D_\text{Ni(Mo)/Ti}<\num{8.5e-5}$, and $D_\text{Cu/Ti(2N6)}<\num{6.0e-5}$ at the $1\sigma$ confidence level, confirming that all samples are depolarizing well below the $\num{e-4}$ level. Given the unknown background of scattered neutrons present in the black configuration and the incomplete polarization by the $^3$He cells, these upper limits are conservative.

\section{Outlook/Conclusion}
The OTB measurements confirmed that all three samples Ni(Mo)/Ti, Cu/Ti(4N5), and Cu/Ti(2N6) showed a depolarization well below $\num{e-4}$, making them suitable for the use in next generation $\beta$ decay experiments like PERC. This is, to the best of our knowledge, the first measurement to reach this precision. From the OTB measurement, a significant difference in the depolarizing properties between the samples cannot be derived. This would require a correction of systematics on the level of $\num{e-5}$ and an improved statistics. Systematics related to the $^3$He spin filters can be reduced by monitoring the cell transmission over time and by measuring the transmission of the completely depolarized cells before exchanging them, allowing to determine the initial $^3$He polarization, the opacity and the relaxation time constant for each installed pair of cells. A neutron supermirror in front of the cells would allow separating the axis of the experiment from the original beam axis, suppressing background from gammas and short wavelength neutrons in the direct beam.\\
Alternatively, a crystal monochromator would further increase the spatial separation from the original beam axis and in addition reduce the uncertainty of the polarization calculation due to the more narrow wavelength band but would also reduce the statistics substantially. Finally, an improved shielding all along the neutron flight path or beam passage in vacuum where possible may further reduce beam-related background that is difficult to assess with closed-shutter measurements. Although such improvements may allow to finally select the best among the three supermirrors in terms of depolarization, the used setup and measuring procedure are adequate for validating the depolarization of samples at the level of below $\num{e-4}$.

\section*{Acknowledgements}
The authors would like to thank D. Berruyer (PF1B, ILL) and the technicians of the central workshop of the TUM physics department for manufacturing essential pieces of the setup, P. Mutti and F. Rey (Instrument Control Group, ILL) for configuring control and readout of the opaque test bench, the technicians Daniel Strobel and Senay Öztürk (Neutron Optics Group, FRM II) for preparing the samples, and the technician Olivier Aguettaz (SuperADAM) for configuring the operation of the cryostat of the second Beryllium filter. Free-of-charge instrument/beam times were offered by the ILL within proposal 1-10-53. The authors also thank R. Cubitt, N.-J. Steinke and M. Bonnaud from the instrument D33 (ILL) for lending the electromagnet and Michael Schulz from the Antares group (FRM II) for lending the first Beryllium filter. The OTB measurement was only possible with the support of D. Jullien and P. Chevalier from the ILL Neutron Optics Group providing the polarized $^3$He cells on a daily basis. Funded by the Deutsche Forschungsgemeinschaft (DFG, German Research Foundation) under Germany’s Excellence Strategy – EXC-2094 – 390783311.

\appendix
\section{Short Wavelength Neutron Contribution}
\label{appfnc}
\begin{figure}[h]
    \centering
		\includegraphics[width=0.48\textwidth]{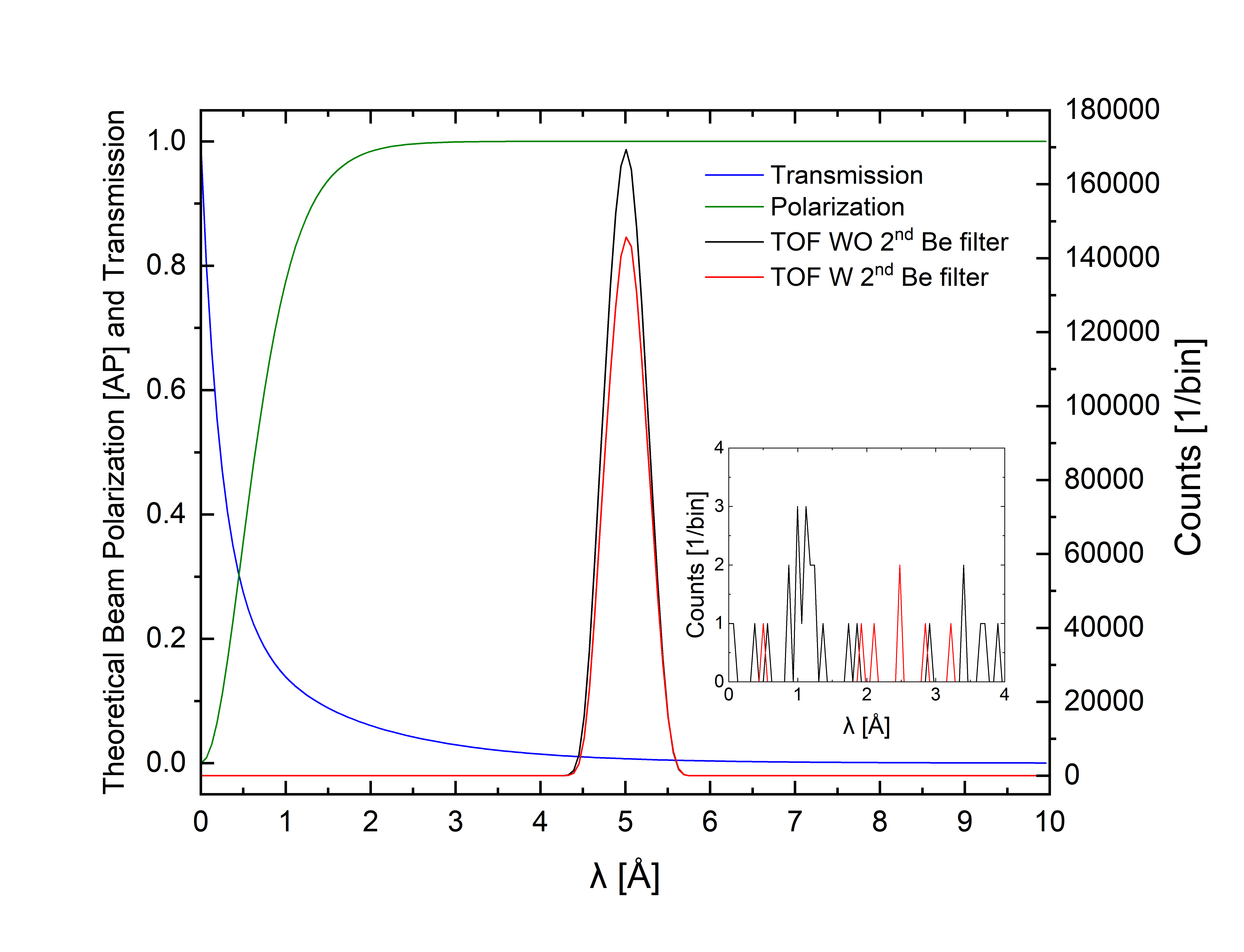}
  \caption{Wavelength spectra of the time of flight measurement at $v_\text{selector} \widehat = \SI{5}{\angstrom}$ with (red line) and without (black line) the Beryllium Filter 2 installed. Also shown is the theoretical probability of transmission (blue line) and degree of polarization (green line) of a pair of identical $^3$He cells with $l=\SI{15}{\centi\metre}$, $P_\text{He}=\SI{79}{\percent}$, and $p=\SI{1.586}{\bar}$ as a function of $\lambda$. The inlet plot is a zoom on the short wavelength neutron part of the spectra, with a clear peak at around $\SI{1}{\angstrom}$ for the spectrum without the Be filter 2. Although this is only a $\num{1.2e-5}$ contribution to the full spectrum, it is enough to significantly worsen $AP$ due to the higher transmission but lower polarization for $\SI{1}{\angstrom}$ neutrons compared to $\SI{5}{\angstrom}$ neutrons.}
  \label{tof}
\end{figure}

During a time of flight study used to verify the selector alignment, the time of flight spectra at a selector speed corresponding to a mean wavelength of $\SI{5}{\angstrom}$ were measured for $\SI{6000}{\second}$ with and without the Beryllium Filter 2 installed, to confirm the short wavelength neutron contribution. The two spectra are shown in figure \ref{tof}, together with the theoretical neutron beam transmission and polarization by two identical $^3$He cells with $p=\SI{1.586}{\bar}$, $P_{\text{He}}=\SI{79}{\percent}$, and $l=\SI{15}{\centi\metre}$. The inlet plot shows a zoom on the lower wavelength range, where an increased rate in short wavelength neutrons of $\lambda\approx\SI{1}{\angstrom}$ in the spectrum without the Beryllium Filter 2 can be observed compared to the spectrum with the Beryllium Filter 2 installed. Although these neutrons contribute only on the order of $\num{e-5}$ compared to the maximum rate at the selected wavelength of $\SI{5}{\angstrom}$, the convolution of each spectrum with the theoretical transmission and polarization of the cells show that the theoretical $AP$ improved from $0.99981(11)$ to $0.999975(19)$ with the use of the Beryllium Filter 2. Note that this calculation includes the single counts outside the peaks at $\SI{1}{\angstrom}$ and $\SI{5}{\angstrom}$, which may be background of the time-of-flight measurement only (e.g. leakage through the chopper disk) that is not relevant for the polarization measurements.

\newpage
\begin{thebibliography}{10}
\expandafter\ifx\csname url\endcsname\relax
  \def\url#1{\texttt{#1}}\fi
\expandafter\ifx\csname urlprefix\endcsname\relax\def\urlprefix{URL }\fi
\expandafter\ifx\csname href\endcsname\relax
  \def\href#1#2{#2} \def\path#1{#1}\fi

\bibitem{MD77}
F.~Mezei, P.~A. Dagleish, Corrigendum and first experimental evidence on
  neutron supermirrors, Commun. Phys. 45 (1977) 41--43.

\bibitem{Sch16}
C.~Schanzer, M.~Schneider, P.~Böni, Neutron optics: Towards applications for
  hot neutrons, J. Phys.: Conf. Ser. 746~(1) (2016) 012024.
\newblock \href {https://doi.org/10.1088/1742-6596/746/1/012024}
  {\path{doi:10.1088/1742-6596/746/1/012024}}.

\bibitem{Erk08}
A.~Erko, M.~Idir, T.~Krist, A.~Michette (Eds.), Modern Developments in X-ray
  and Neutron Optics, Springer Series in Optical Sciences, Springer, 2008.

\bibitem{Hol+22}
A.~Hollering, N.~Rebrova, C.~Klauser, T.~Lauer, B.~Märkisch, U.~Schmidt, A
  non-depolarizing {CuTi} neutron supermirror guide for {PERC}, Nucl. Instr.
  Meth. A 1032 (2022) 166634.
\newblock \href {https://doi.org/10.1016/j.nima.2022.166634}
  {\path{doi:10.1016/j.nima.2022.166634}}.

\bibitem{Dub08}
D.~Dubbers, H.~Abele, S.~Baeßler, B.~Märkisch, M.~Schumann, T.~Soldner,
  O.~Zimmer, A clean, bright, and versatile source of neutron decay products,
  Nucl. Instr. Meth. A 596~(2) (2008) 238--247.
\newblock \href {https://doi.org/10.1016/j.nima.2008.07.157}
  {\path{doi:10.1016/j.nima.2008.07.157}}.

\bibitem{Wan+19}
X.~Wang, C.~Ziener, H.~Abele, S.~Bodmaier, D.~Dubbers, J.~Erhart, A.~Hollering,
  E.~Jericha, J.~Klenke, H.~Fillunger, W.~Heil, C.~Klauser, G.~Konrad,
  M.~Lamparth, T.~Lauer, M.~Klopf, R.~Maix, B.~Märkisch, W.~Mach, H.~Mest,
  D.~Moser, A.~Pethoukov, L.~Raffelt, N.~Rebrova, C.~Roick, H.~Saul,
  U.~Schmidt, T.~Soldner, R.~Virot, O.~Zimmer, {(PERC collaboration)}, Design
  of the magnet system of the neutron decay facility {PERC}, EPJ Web of
  Conferences 219 (2019) 04007.
\newblock \href {https://doi.org/10.1051/epjconf/201921904007}
  {\path{doi:10.1051/epjconf/201921904007}}.

\bibitem{Cab86}
J.~W. Cable, M.~R. Khan, G.~P. Felcher, I.~K. Schuller, Macromagnetism and
  micromagnetism in {Ni}-{Mo} metallic superlattices, Phys. Rev. B 34~(3)
  (1986) 1643--1649.
\newblock \href {https://doi.org/10.1103/PhysRevB.34.1643}
  {\path{doi:10.1103/PhysRevB.34.1643}}.

\bibitem{Pad04}
J.~Padiyath, J.~Stahn, P.~Allenspach, M.~Horisberger, P.~Böni, Influence of
  {Mo} in the {Ni} sublayers on the magnetization of {Ni}/{Ti} neutron
  supermirrors, Physica B: Condensed Matter 350~(1-3) (2004) E237--E240.
\newblock \href {https://doi.org/10.1016/j.physb.2004.03.059}
  {\path{doi:10.1016/j.physb.2004.03.059}}.

\bibitem{Kov08}
R.~Kovács-Mezei, T.~Krist, Z.~Révay, Non-magnetic supermirrors produced at
  {Mirrotron} {Ltd}., Nucl. Instr. Meth. A 586~(1) (2008) 51--54.
\newblock \href {https://doi.org/10.1016/j.nima.2007.11.034}
  {\path{doi:10.1016/j.nima.2007.11.034}}.

\bibitem{KAN97}
F.~Khan, M.~Asgar, P.~Nordblad, Magnetization and magnetocrystalline anisotropy
  of {Ni}-{Mo} single crystal alloys, Journal of Magnetism and Magnetic
  Materials 174~(1-2) (1997) 121--126.
\newblock \href {https://doi.org/10.1016/S0304-8853(97)00182-0}
  {\path{doi:10.1016/S0304-8853(97)00182-0}}.

\bibitem{Sch99}
A.~Schebetov, A.~Kovalev, B.~Peskov, N.~Pleshanov, V.~Pusenkov,
  P.~Schubert-Bischoff, G.~Shmelev, Z.~Soroko, V.~Syromyatnikov, V.~Ul'yanov,
  A.~Zaitsev, Multi-channel neutron guides of {PNPI}: results of neutron and
  {X}-ray reflectometry tests, Nucl. Instr. Meth. A 432~(2-3) (1999) 214--226.
\newblock \href {https://doi.org/10.1016/S0168-9002(99)00480-5}
  {\path{doi:10.1016/S0168-9002(99)00480-5}}.

\bibitem{Reb14}
N.~Rebrova, Developement of a non-depolarizing neutron guide for {PERC}, {PhD},
  Ruprecht-Karls-Universität, Heidelberg (May 2014).

\bibitem{Bon17}
V.~Bondar, S.~Chesnevskaya, M.~Daum, B.~Franke, P.~Geltenbort, L.~Göltl,
  E.~Gutsmiedl, J.~Karch, M.~Kasprzak, G.~Kessler, K.~Kirch, H.-C. Koch,
  A.~Kraft, T.~Lauer, B.~Lauss, E.~Pierre, G.~Pignol, D.~Reggiani,
  P.~Schmidt-Wellenburg, Y.~Sobolev, T.~Zechlau, G.~Zsigmond, Losses and
  depolarization of ultracold neutrons on neutron guide and storage materials,
  Phys. Rev. C 96~(3) (2017) 035205.
\newblock \href {https://doi.org/10.1103/PhysRevC.96.035205}
  {\path{doi:10.1103/PhysRevC.96.035205}}.

\bibitem{Ple94}
N.~Pleshanov, V.~Pusenkov, A.~Schebetov, B.~Peskov, G.~Shmelev, E.~Siber,
  Z.~Soroko, On the use of specular neutron reflection in the study of
  roughness and interdiffusion in thin-film structures, Physica B: Condensed
  Matter 198~(1-3) (1994) 27--32.
\newblock \href {https://doi.org/10.1016/0921-4526(94)90119-8}
  {\path{doi:10.1016/0921-4526(94)90119-8}}.

\bibitem{Kla12}
C.~Klauser, J.~Chastagnier, D.~Jullien, A.~Petoukhov, T.~Soldner, High
  precision depolarisation measurements with an opaque test bench, Journal of
  Physics: Conference Series 340 (2012) 012011.
\newblock \href {https://doi.org/10.1088/1742-6596/340/1/012011}
  {\path{doi:10.1088/1742-6596/340/1/012011}}.

\bibitem{Kla13}
C.~Klauser, High {Precision} {Neutron} {Polarization} {For} {PERC}, {PhD},
  Technische Universität Wien, Vienna (Oct. 2013).

\bibitem{Gom24}
J.~M. Gómez-Guzmán, M.~Opel, T.~Veres, P.~Link, L.~Bottyán, Structural,
  electrical and magnetic properties of reactively {DC} sputtered {Cu} and {Ti}
  thin films. {Application} to {Cu}/{Ti} neutron supermirrors for low spin-flip
  applications, Nucl. Instr. Meth. A 1059 (2024) 169005.
\newblock \href {https://doi.org/10.1016/j.nima.2023.169005}
  {\path{doi:10.1016/j.nima.2023.169005}}.

\bibitem{Anton13}
A.~Devishvili, K.~Zhernenkov, A.~Dennison, B.~Toperverg, B.~Hjörvarsson,
  H.~Zabel, Super{ADAM}: {U}pgraded polarized neutron reflectometer at the
  {I}nstitut {L}aue-{L}angevin, The {R}eview of {S}cientific {I}nstruments 84
  (2013) 025112.
\newblock \href {https://doi.org/10.1063/1.4790717}
  {\path{doi:10.1063/1.4790717}}.

\bibitem{SADAM}
{Institut Laue-Langevin},
  \href{https://www.ill.eu/users/instruments/instruments-list/superadam/}{{SuperADAM}},
  accessed: 2024-08-22 (2024).
\newline\urlprefix\url{https://www.ill.eu/users/instruments/instruments-list/superadam/}

\bibitem{Kla16}
C.~Klauser, T.~Bigault, P.~Böni, P.~Courtois, A.~Devishvili, N.~Rebrova,
  M.~Schneider, T.~Soldner, Depolarization in polarizing supermirrors, Nucl.
  Instr. Meth. A 840 (2016) 181--185.
\newblock \href {https://doi.org/10.1016/j.nima.2016.09.056}
  {\path{doi:10.1016/j.nima.2016.09.056}}.

\bibitem{Pf1b}
{Institut Laue-Langevin},
  \href{https://www.ill.eu/users/instruments/instruments-list/pf1b/}{{PF1B}},
  accessed: 2024-08-21 (2024).
\newline\urlprefix\url{https://www.ill.eu/users/instruments/instruments-list/pf1b/}

\bibitem{Abe06}
H.~Abele, D.~Dubbers, H.~Häse, M.~Klein, A.~Knöpfler, M.~Kreuz, T.~Lauer,
  B.~Märkisch, D.~Mund, V.~Nesvizhevsky, A.~Petoukhov, C.~Schmidt,
  M.~Schumann, T.~Soldner, Characterization of a ballistic supermirror neutron
  guide, Nucl. Instr. Meth. A 562~(1) (2006) 407--417.
\newblock \href {https://doi.org/10.1016/j.nima.2006.03.020}
  {\path{doi:10.1016/j.nima.2006.03.020}}.

\bibitem{Dat23}
J.~M. Gómez-Guzmán, K.~Bernert, A.~Devishvili, C.~Klauser, B.~Märkisch,
  U.~Schmidt, T.~Soldner, Depolarisation of {Cu/Ti} supermirrors.
\newblock \href {https://doi.org/10.5291/ILL-DATA.1-10-53}
  {\path{doi:10.5291/ILL-DATA.1-10-53}}.

\bibitem{Mas}
S.~Masalovich, Analysis and design of multilayer structures for neutron
  monochromators and supermirrors, Nucl. Instr. Meth. A 722 (2013) 71--81.
\newblock \href {https://doi.org/10.1016/j.nima.2013.04.051}
  {\path{doi:10.1016/j.nima.2013.04.051}}.

\bibitem{Super}
M.~Wolff, F.~Radu, A.~Petoukhov, H.~Humblot, D.~Jullien, K.~H. Andersen,
  H.~Zabel, Scientific reviews $^3\text{{He}}$ spin filter at the {Institut
  Laue-Langevin}: Polarization analysis of diffuse scattering, Neutron News
  17~(2) (2006) 26--29.
\newblock \href {https://doi.org/10.1080/10448630600668761}
  {\path{doi:10.1080/10448630600668761}}.

\bibitem{Wel}
R.~J. Welbourn, C.~Truscott, M.~A. Skoda, A.~Zarbakhsh, S.~Clarke, Corrosion
  and inhibition of copper in hydrocarbon solution on a molecular level
  investigated using neutron reflectometry and {XPS}, Corrosion Science 115
  (2017) 68--77.
\newblock \href {https://doi.org/https://doi.org/10.1016/j.corsci.2016.11.010}
  {\path{doi:https://doi.org/10.1016/j.corsci.2016.11.010}}.

\bibitem{Pet06}
A.~Petoukhov, V.~Guillard, K.~Andersen, E.~Bourgeat-Lami, R.~Chung, H.~Humblot,
  D.~Jullien, E.~Lelievre-Berna, T.~Soldner, F.~Tasset, M.~Thomas, Compact
  magnetostatic cavity for polarised $^3\text{{He}}$ neutron spin filter cells,
  Nucl. Instr. Meth. A 560~(2) (2006) 480--484.
\newblock \href {https://doi.org/10.1016/j.nima.2005.12.247}
  {\path{doi:10.1016/j.nima.2005.12.247}}.

\bibitem{Cou88}
K.~Coulter, A.~McDonald, W.~Happer, T.~Chupp, M.~Wagshul, Neutron polarization
  with polarized $^3\text{{He}}$, Nucl. Instr. Meth. A 270~(1) (1988) 90--94.
\newblock \href {https://doi.org/10.1016/0168-9002(88)90013-7}
  {\path{doi:10.1016/0168-9002(88)90013-7}}.

\bibitem{Fri89}
H.~Friedrich, V.~Wagner, P.~Wille, A high-performance neutron velocity
  selector, Physica B: Condensed Matter 156-157 (1989) 547--549.
\newblock \href {https://doi.org/10.1016/0921-4526(89)90727-8}
  {\path{doi:10.1016/0921-4526(89)90727-8}}.

\bibitem{gehrels}
N.~Gehrels, Confidence limits for small numbers of events in astrophysical
  data, the Astrophysical Journal 303 (1986) 336--346.
\newblock \href {https://doi.org/10.1086/164079} {\path{doi:10.1086/164079}}.

\bibitem{TYR2}
{Institut Laue-Langevin},
  \href{https://www.ill.eu/users/instruments/instruments-list/tyrex-2/}{{TYREX}-2},
  accessed: 2024-06-14 (2024).
\newline\urlprefix\url{https://www.ill.eu/users/instruments/instruments-list/tyrex-2/}

\end{thebibliography}
\end{document}